# The quantum theory of time: a calculus for q-numbers

By Samuel Kuypers[1]
May 2021

**Abstract**

In quantum theory, physical systems are usually assumed to evolve relative to a c-number time. This c-number time is unphysical and has turned out to be unnecessary for explaining dynamics: in the timeless approach to quantum theory developed by Page & Wootters (1983), subsystems of a stationary universe can instead evolve relative to a 'clock', which is a quantum system with a q-number time observable. Page & Wootters formulated their construction in the Schrödinger picture, which left open the possibility that the c-number time still plays an explanatory role in the Heisenberg picture. I formulate their construction in the Heisenberg picture and demonstrate how to eliminate c-number time from that picture, too. When the Page–Wootters construction is formulated in the Heisenberg picture, the descriptors of physical systems are functions of the clock's q-number time, and derivatives with respect to this q-number time can be defined in terms of the clock's algebra of observables, which results in a calculus for q-numbers.

## 1 Introduction

It is a principle of quantum theory that all descriptors[2] of physical systems are q-numbers[3]. For example, the momentum and position of a particle are canonically conjugate q-numbers; the descriptors of a qubit adhere to the Pauli algebra; and the descriptors of fermions are Grassmann operators. Due to its significance, I shall call this the *fundamental principle of quantum theory.*

---

[1] The Clarendon Laboratory, University of Oxford, Oxford OX1 3PU, UK

[2] The descriptors of a physical system are mathematical objects – *e.g.* tensors, vectors, real numbers, integers – that represents the state of the system.

[3] Q-numbers are numbers that satisfy a non-commutative algebra; they are conventionally represented by operators or Hermitian matrices on a Hilbert space. C-numbers satisfy a commutative algebra – examples complex numbers are c-numbers. In quantum theory, a system's state is specified by a set of q-numbers and a state vector. For instance, a qubit has an $x$-, $y$-, and $z$-observable (which are q-numbers), and the state of the qubit is represented by those q-numbers and their expectation values.

Because of this fundamental principle, time has a peculiar status within quantum theory: time is a c-number, and so it cannot be the descriptor of a physical system. This c-number time is, instead, an unphysical backdrop against which quantum systems can change. Hence, c-number time is all but identical to Newton's (1687) absolute time, which

> *'of itself, and from its own nature flows equably without regard to anything external, and by another name is called duration…'*

Put differently, c-number time 'flows' independently of the behaviour of quantum systems, and it should not be confused with clock time.

To make clear the distinction between clock time and c-number time, consider a hypothetical situation in which all physical systems in the universe suddenly 'freeze' relative to c-number time.[4] Clocks are physical systems and would be stationary during this 'freeze', which makes the 'freeze' and its duration completely undetectable to any observer in the universe. Thus, c-number time must be just as unmeasurable as the duration of the 'freeze'.

C-number time 'flows without regard to anything external' and is, therefore, unphysical. However, clock time is physical: it is a descriptor of a clock, which is a physical system. Clock time is also measurable: it can be straightforwardly measured, by the usual methods, on the clock. Hence, in classical physics it is possible to make time manifestly physical by replacing Newton's absolute time with clock time. For example, Barbour (2009) has shown that a universe consisting solely of classical systems that interact with one another through Newtonian gravity can give rise to clocks, and that the dynamical laws of such a universe can be formulated entirely in terms of clock time.

Similarly, Page & Wootters (*op. cit.*) replaced quantum theory's c-number time with clock time. They did this by proposing a quantum model of a completely stationary universe – *i.e.* its universal state-vector $|\Psi_S\rangle$ is an eigenstate of the universe's Hamiltonian so that no measurable quantity of this universe will depend on c-number time. Hence, it is as if the c-number time is not there at all. Within this stationary universe, systems can instead change relative to one another. In particular, they can change relative to a quantum clock, which has an associated q-number known as the *time observable* that completely replaces the role of c-number time. So, the Page–Wootters construction resolves the conflict between c-number time and the fundamental principle of quantum theory – at least in the Schrödinger picture.

[4] This thought experiment is based on Shoemaker's (1969) discussion of Newton's absolute time.

However, it is not clear that the Page–Wootters construction eliminates the need for c-number time from the Heisenberg picture. Although the Heisenberg and Schrödinger pictures are mathematically equivalent, they are conceptually different: in the Heisenberg picture, the time-varying entities are the system's q-number descriptors, which affect each other only via local interactions, whereas the Schrödinger-picture state vector is non-local. And even when a system's Schrödinger-picture state vector is time-invariant, that system's Heisenberg descriptors can still depend on c-number time (see §3). If c-number time plays an explanatory role in the Heisenberg-picture Page–Wootters construction, the conflict with the fundamental principle of quantum theory would not have been resolved after all, at least not in that picture.

Suppose that the Page–Wootters construction cannot be adequately formulated in the Heisenberg picture – *e.g.* perhaps the Heisenberg picture cannot validly describe the entire universe. In that case, the Schrödinger picture would be the more fundamental description of quantum theory. This would be problematic since there are good reasons to regard the Heisenberg picture as not less but *more* fundamental than the Schrödinger picture. For instance, unlike the Schrödinger picture, the Heisenberg picture is a local description of quantum systems (Deutsch & Hayden 2000, Raymond-Robichaud 2021).[5] If the Heisenberg picture is the more fundamental description of quantum theory, it should admit a formulation of the Page–Wootters construction that is independent of c-number time.

In this paper, I demonstrate that the Page–Wootters construction eliminates the need for c-number time in both the Schrödinger and the Heisenberg picture. When the Page–Wootters construction is formulated in the Heisenberg picture, the descriptors of physical system will turn out to be functions of the clock's q-number time observable. Moreover, one can define derivatives with respect to q-number time in terms of the clock's algebra of observables, so that it is possible to refer solely to q-number time in formulating equations of motion. The theory of q-number functions and their derivatives is what I will call the *q-number calculus*.[6]

[5] That is to say that only the Heisenberg picture satisfies Einsteinian locality, which requires that 'the real factual situation of the system $S_2$ is independent of what is done with the $S_1$, which is spatially separated from the former' (Einstein 1970, p. 85).

## 2 Page–Wootters in the Schrödinger picture

The Page–Wootters construction describes a block universe $\mathfrak{U}$.[7] In the Schrödinger picture, this model universe is stationary because its universal state-vector $|\Psi_S\rangle$ is an eigenstate of the universe's Hamiltonian $\mathcal{H}_S$,

$$\hat{\mathcal{H}}_S|\Psi_S\rangle = 0. \tag{2.1}$$

Here and throughout, I use the subscript $S$ to indicate that a q-number or a vector is part of the Schrödinger picture. Due to (2.1), it now appears as if this block universe is completely timeless, but subsystems of this universe can nonetheless evolve relative to one another. In particular, they can evolve relative to a clock.

In the Page–Wootters construction, a clock $\mathfrak{C}$ is a quantum system with the constitutive property that it possesses a pair of descriptors $\hat{h}_S$ and $\hat{t}_S$ that are canonically conjugate

$$[\hat{t}_S, \hat{h}_S] = i\hat{1}. \tag{2.2}$$

$\hbar$ has been set equal to 1 throughout this paper. The q-numbers $\hat{h}_S$ and $\hat{t}_S$ live on a Hilbert space $\boldsymbol{H}_C$, and $\hat{1}$ denotes the unit observable on that space. $\hat{t}_S$ is the clock's *time observable*, so named because the eigenstates of $\hat{t}_S$ represent different times – *i.e.* the eigenstate $|t\rangle$ of $\hat{t}_S$ with eigenvalue $t$ has the physical meaning 'it is clock time $t$'.

[6] Perhaps the first person to conceive of using the Heisenberg picture to formulate a calculus for q-numbers was Dirac (1926). I shall expand on Dirac's ideas in later sections by clarifying the notion of a q-number function and its derivatives.

[7] An early, if not the earliest, iteration of the block universe theory was invented over two millennia ago, around 500 BC, by the pre-Socratic Greek philosopher Parmenides. Parmenides expounded that the universe is fundamentally unchanging; that all movement is only apparent; and that through the use of reason one can discover that change is an illusion (Popper 2012). Parmenides' block universe is remarkably similar to the Page–Wootters construction, as each describes a stationary universe in which time and dynamics are emergent.

The q-number $\hat{h}_S$ generates translations of $\hat{t}_S$: by using the algebraic relation (2.2), one can readily verify that an eigenstate $|t\rangle$ of $\hat{t}_S$ is shifted by a c-number factor $\lambda$ when it is multiplied by the translation operator $e^{-i\hat{h}_S\lambda}$, *i.e.*

$$e^{-i\hat{h}_S\lambda}|t\rangle = |t+\lambda\rangle, \tag{2.3}$$

and if $\lambda$ is taken to be infinitesimally small, it follows from (2.3) that

$$\hat{h}_S|t\rangle = i\frac{d}{dt}|t\rangle. \tag{2.4}$$

Certain properties of the clock are invariant for all its states. One such property, the constitutive algebra (2.2), has been shown already. A second invariant property is the spectrum of eigenvalues of $\hat{t}_S$, denoted $\mathrm{Sp}(\hat{t}_S)$. The spectrum $\mathrm{Sp}(\hat{t}_S)$ is equal to $\mathbb{R}$ because (2.3) implies that, if some real number $t$ is an eigenvalue of $\hat{t}_S$, then $t+\lambda$ must also be an eigenvalue of $\hat{t}_S$, for all $\lambda \in \mathbb{R}$. In a similar vein, the spectrum of eigenvalues of $\hat{h}_S$ can be shown to be equal to $\mathbb{R}$ as well.

$\mathfrak{C}$ functions as an ideal clock if the 'rest' of the universe evolves according to the Schrödinger equation relative to $\mathfrak{C}$, where the 'rest', denoted $\mathfrak{R}$, constitutes the set of quantum systems in $\mathfrak{U}$ apart from the clock. As will be shown below, the state of $\mathfrak{R}$ evolves according to the Schrödinger equation relative to clock time if $\mathfrak{C}$ and $\mathfrak{R}$ are dynamically isolated from one another. Hence, Hamiltonian of the universe, denoted $\hat{\mathcal{H}}_S$, should contain no interaction terms, so that

$$\hat{\mathcal{H}}_S = \hat{H}_S + \hat{h}_S. \tag{2.5}$$

Here the Hamiltonian of the clock is $\hat{h}_S$, and $\hat{H}_S$ is the Hamiltonian of $\mathfrak{R}$.

One obtains the state of $\mathfrak{R}$ relative to the clock being in the state 'it is clock time $t$' by taking the partial inner-product of the universal state-vector $|\Psi_S\rangle$ and the eigenstate $|t\rangle$ of $\hat{t}_S$,

$$|\psi(t)\rangle \stackrel{\text{def}}{=} \langle t|\Psi_S\rangle\rangle. \quad (2.6)$$

The $\rangle\rangle$ show, where necessary, that $|\Psi_S\rangle$ resides on a larger Hilbert space than $|t\rangle$. By using (2.6), (2.5), (2.4), and (2.1), one can now deduce that $|\psi(t)\rangle$ obeys the Schrödinger equation (Wootters 1984),

$$i\frac{d}{dt}|\psi(t)\rangle = i\frac{d}{dt}\langle t|\Psi_S\rangle\rangle = -\langle t|\hat{h}_S|\Psi_S\rangle\rangle = \langle t|\hat{H}_S|\Psi_S\rangle\rangle = \hat{H}_S|\psi(t)\rangle.$$

Thus, the eigenvalues of $\hat{t}_S$ correspond to the time-parameter of the 'rest' of the universe, and therefore one can formulate dynamics in terms of clock time, despite the universe being stationary with respect to c-number time.

If a moment such as $|\psi(t)\rangle|t\rangle$ is likened to a snapshot, the universal state-vector $|\Psi_S\rangle$ is like a photo album containing the totality of such snapshots. To put this less metaphorically and more mathematically, the universal state-vector is the integral of the relative states over all possible clock times

$$|\Psi_S\rangle = \int_{-\infty}^{\infty} |\psi(t)\rangle|t\rangle dt.$$

This way of formulating the Page–Wootters construction shows its deep connection to Everett's relative-state formalism (1973): the state $|\psi(t)\rangle|t\rangle$ is a relative state, representing $|\Psi_S\rangle$ relative to the clock being in the state $|t\rangle$. Due to the role that Everett's relative-state formalism plays in the Page–Wootters construction, different times are equivalent to different Everett universes (Deutsch 1997).[8]

[8] For $\mathfrak{C}$ to be an ideal clock, there can be no interactions between it and the rest of the universe. Hence, if one were to measure $\mathfrak{C}$, it would cease to be an ideal clock. So, time can only be measured imperfectly in the Page–Wootters construction. This does not affect other features of the construction – for instance, the universal state vector can nonetheless be decomposed into relative states (relative to clock time) because the clock and the rest of the universe need only be entangled for that.

Page & Wootters' achievement is that they constructed a quantum theory of clocks, and that in their block universe, all change must be change relative to such clocks. In this way, the unphysical c-number time is made completely unnecessary for explaining change and dynamics, which solves the conflict between c-number time and the fundamental principle of quantum theory.

## 3 Q-number time in the Heisenberg picture

Moving between the Heisenberg and Schrödinger pictures is typically a straightforward procedure that consists of applying unitaries to a system's descriptors and state vector. However, this procedure is not invariably straightforward: because of conceptual difficulties, Everett's relative-state formalism has only recently been translated to the Heisenberg picture (Kuypers & Deutsch 2021). And since Everett's relative-state formalism is essential to formulating the Page–Wootters construction, similar issues also appear in translating that construction to the Heisenberg picture.

One such issue is that that the Heisenberg descriptors might still depend on c-number time. Consider, for instance, a qubit $\mathfrak{Q}$ in the Heisenberg picture. Following Deutsch & Hayden (*op. cit.*), the state of the qubit at c-number time $\lambda$ is represented by a fixed Heisenberg state and a triple of q-number descriptors,

$$\hat{\boldsymbol{q}}(\lambda) = \left(\hat{q}_x(\lambda),\ \hat{q}_y(\lambda),\ \hat{q}_z(\lambda)\right),$$

that satisfy the Pauli algebra at all times $\lambda$

$$\left.\begin{aligned} \hat{q}_i(\lambda)\hat{q}_j(\lambda) &= \delta_{ij}\hat{1} + i\epsilon_{ij}{}^{k}\hat{q}_k(\lambda) \\ \left(\hat{q}_i(\lambda)\right)^\dagger &= \hat{q}_i(\lambda) \end{aligned}\right\} \qquad (i,j,k \in \{x,y,z\}).$$

Here † denotes Hermitian conjugation. Throughout this paper, I shall use Einstein's summation convention by summing over indices that appear twice in a product, so in the expression above, $k$ is summed over all its possible values $\{x,y,z\}$. Furthermore, if $\mathfrak{Q}$ is an isolated system, its descriptors $\hat{\boldsymbol{q}}(\lambda)$ evolve unitarily and satisfy the Heisenberg equation of motion

$$\frac{d\hat{\boldsymbol{q}}(\lambda)}{d\lambda} = i[\hat{H}, \hat{\boldsymbol{q}}(\lambda)], \tag{3.1}$$

where $\hat{H}$ is the qubit's Hamiltonian.

In the Schrödinger-picture Page–Wootters construction, $\lambda$ is irrelevant to the system's description because all measurable quantities are assumed to be independent of $\lambda$. Yet, in the Heisenberg picture, even when the expectation values of the descriptors $\hat{\boldsymbol{q}}(\lambda)$ do not depend on $\lambda$, those descriptors will in general depend on c-number time. For instance, the equation of motion (3.1) will be formulated in terms of $\lambda$, and the descriptors will be functions of $\lambda$, regardless of their expectation values. Thus, c-number time apparently plays a more fundamental role in the Heisenberg picture than it does in the Schrödinger picture.

But it need not play any role. The Heisenberg equation of motion of the Page–Wootters universe can be formulated entirely in terms of q-number time. And because such a formulation is possible, c-number can be all but expunged from the Heisenberg picture. To formulate an equation of motion in terms of the q-number time $\hat{t}$, I shall first need to describe, in the Heisenberg picture, what a function of $\hat{t}$ is and how to define derivatives of such functions. Functions of $\hat{t}$ can be defined in terms of the set of *possible states* that a clock can be in.

## 3.1 Possible states

Physical systems are defined by a set of possible states – 'possible' in the counterfactual sense that these are the states that the system *could be in*, regardless of the dynamics of the system. For example, consider a qubit whose state is represented by a Heisenberg state $|\Psi\rangle$ and a triple of time-invariant descriptors $\hat{\boldsymbol{q}}$ that satisfy the Pauli algebra. For fixed $|\Psi\rangle$, this qubit's set of possible states can be defined as follows: $\mathbb{P} \stackrel{\text{def}}{=} \{U^\dagger \hat{\boldsymbol{q}} U : U^\dagger U = U U^\dagger = \hat{1}\}$, where $U$ is a function solely of $\hat{\boldsymbol{q}}$.

Physical systems also have a *history*, which is a one-parameter family of states uniquely specified by the system's equations of motion and initial conditions. For a fixed Heisenberg state, the qubit's history $\mathbb{H}$ is a set of triples $\hat{\boldsymbol{q}}(\lambda) \in \mathbb{P}$ that depend on a c-number time $\lambda$ and adhere to the qubit's equation of motion. The parameterised $\hat{\boldsymbol{q}}(\lambda)$ must also satisfy an initial condition, say $\hat{\boldsymbol{q}}(0) = \hat{\boldsymbol{q}}_0 \in \mathbb{P}$, so that $\mathbb{H}$ is a parameterised curve on $\mathbb{P}$.

There is a key difference between the time-dependent $\hat{\boldsymbol{q}}(\lambda)$ in $\mathbb{H}$ and the more fundamental descriptors $\hat{\boldsymbol{q}}$ in $\mathbb{P}$: the triple $\hat{\boldsymbol{q}}$ is part of the counterfactual set of states that the system *could be in*, independently of which states the system *will actually be in* according to its dynamical laws and initial conditions. Hence, the triple $\hat{\boldsymbol{q}}$ and Heisenberg state $|\Psi\rangle$ represent a more fundamental, timeless notion of state. It is this timeless notion of state that I will refer to throughout the following sections, unless stated otherwise.

## 3.2 The possible states of the clock

In the Heisenberg picture, a possible state of the clock $\mathfrak{C}$ is fully specified by a Heisenberg state $|\Psi\rangle$ and two Hermitian descriptors $\hat{t}$ and $\hat{h}$ that are canonically conjugate, as in (2.2). These q-numbers live on a Hilbert space $\boldsymbol{H}_C$. Just like in the Schrödinger picture, the spectrum of eigenvalues of $\hat{t}$ is $\mathrm{Sp}(\hat{t}) = \mathbb{R}$. The individual eigenvalues of $\hat{t}$ can be analysed through a projection operator $\hat{\Pi}_t(\hat{t})$, which has the following property

$$\hat{t}\hat{\Pi}_t(\hat{t}) = t\hat{\Pi}_t(\hat{t}) \qquad \Big(\text{for all } t \in \mathrm{Sp}(\hat{t})\Big). \tag{3.2}$$

These projectors sum to unity, making the set $\{\hat{\Pi}_t(\hat{t}) : t \in \mathrm{Sp}(\hat{t})\}$ a projection-valued measure, in terms of which the q-number $\hat{t}$ has a spectral decomposition

$$\hat{t} = \int_{-\infty}^{\infty} t\hat{\Pi}_t(\hat{t})dt.$$

$\hat{t}$ and $\hat{h}$, and all other q-number descriptors of the clock, have expectation values that are determined by the Heisenberg state $|\Psi\rangle$.[9] The expectation value of an arbitrary descriptor $\hat{A}$ shall be denoted as $\langle\hat{A}\rangle \stackrel{\text{def}}{=} \langle\Psi|\hat{A}|\Psi\rangle$. Furthermore, when this descriptor has the property that

[9] Other observables of the clock are for instance $\hat{t}^2$ and $\hat{h}^2$, and their expectation values can in general not be deduced from those of $\hat{t}$ and $\hat{h}$.

$\langle \hat{A} \rangle^2 = \langle \hat{A}^2 \rangle$, the Heisenberg state $|\Psi\rangle$ is an eigenstate of $\hat{A}$ with eigenvalue $\langle \hat{A} \rangle$, and $\hat{A}$ is then said to be *sharp* with that eigenvalue.

A state of the clock for which $\hat{t}$ is sharp with the value $t$ has the unambiguous physical meaning 'it is clock time $t$', and the clock is then said to be in a *clock state* since it is a state of the clock that represents a definite time. When $\hat{t}$ is non-sharp, the clock is in a superposition of clock states; in which case there is no fact about which time the clock represents. Yet, even when $\hat{t}$ is non-sharp, there is a fact about the time that the clock represents in a *relative state*. For instance, when $\hat{t}$ is non-sharp, one can consider other systems in the Page–Wootters universe relative to the clock being in the state 'it is clock time $t$' by, for example, projecting the Heisenberg state onto a *relative Heisenberg state* $|\Psi_t\rangle$, which is defined as follows:

$$|\Psi_t\rangle \stackrel{\text{def}}{=} \frac{\widehat{\Pi}_t(\hat{t})|\Psi\rangle}{\langle \widehat{\Pi}_t(\hat{t}) \rangle},$$

where the factor $\langle \widehat{\Pi}_t(\hat{t}) \rangle$ in the denominator is assumed to be non-zero and ensures that the relative Heisenberg state is normalised (Kuypers & Deutsch *op. cit.*). The relative Heisenberg state $|\Psi_t\rangle$ is an eigenstate of $\hat{t}$ with eigenvalue $t$, so with respect to $|\Psi_t\rangle$, the clock will read 'it is clock time $t$'. Consequently, even if $\hat{t}$ is non-sharp and there is no fact about what time the clock represents with respect to the absolute state of the system, the clock can represent a definite time within a relative state.

### 3.3 Q-number functions

Because of the fundamental principle of quantum theory, c-number time should play no role in quantum theory. Therefore, a system's equations of motion should be formulated in terms of a q-number time $\hat{t}$, and the solutions to such an equation of motion must be a function of $\hat{t}$. Here I define what a function of $\hat{t}$ is.

Note that $\hat{t}$ has neither a largest nor a smallest eigenvalue on the full Hilbert space $\boldsymbol{H_C}$: it is an unbounded q-number. Its unboundedness results in conceptual difficulties that I will discuss in more detail in §5. What is relevant for the current discussion is that, because $\hat{t}$ is unbounded, it is typically difficult to define a well-behaved function in terms of $\hat{t}$. It is convenient to rely instead on the operator $e^{i\omega\hat{t}}$, where $\omega$ is a real number with units of frequency so that the product $\omega\hat{t}$ is a dimensionless q-number. The value of $\omega$ will be determined by the system's equations of motion – *i.e.* it is determined by the system's Hamiltonian (see §4). Importantly, the q-number $e^{i\omega\hat{t}}$ is bounded because its norm is unity. So, I shall here study q-number functions of the form

$$f(\omega\hat{t}) \stackrel{\text{def}}{=} \sum_{n=-N}^{N} a_n e^{in\omega\hat{t}} , \qquad (3.3)$$

where $N$ is an integer. Such functions are finite sums of the bounded q-number $e^{i\omega\hat{t}}$ and are therefore automatically bounded. They are also periodic.[10]

Furthermore, to guarantee that $f(\omega\hat{t})$ is Hermitian, assume that its coefficients $\{a_n\}$ are complex numbers with the property that $a_n^\dagger = a_{-n}$ for $-N \leq n \leq N$. By virtue of being Hermitian, $f(\omega\hat{t})$ has a spectrum of eigenvalues, denoted $\mathrm{Sp}\big(f(\omega\hat{t})\big)$.

(3.3) does not represent the most general possible function of $\hat{t}$. For instance, there are also convergent series in $e^{i\omega\hat{t}}$, which are not included in that definition (see *e.g.* Kreyszig 1978). But for the purposes of this paper, sums of the form (3.3) are sufficiently general since such q-number functions can be used to study the evolution of finite systems – as I shall show in §4 the evolution of a qubit can, for example, be described in terms of the q-number functions $\cos(\omega\hat{t}) \stackrel{\text{def}}{=} \frac{1}{2}\big(e^{-i\omega\hat{t}} + e^{i\omega\hat{t}}\big)$ and $\sin(\omega\hat{t}) \stackrel{\text{def}}{=} \frac{i}{2}\big(e^{-i\omega\hat{t}} - e^{i\omega\hat{t}}\big)$.

The descriptors of physical systems appear to be functions of c-number time, yet fundamentally all time-dependent functions should be functions of $\hat{t}$. If that is so, how can one account for the appearance of c-number functions? The solution to this problem is as follows: consider that the spectral decomposition of an arbitrary function $f(\omega\hat{t})$ is

$$f(\omega\hat{t}) = \int_{-\infty}^{\infty} f(\omega t)\hat{\Pi}_t(\hat{t})dt. \tag{3.4}$$

where $f(\omega t)$ is identical to the sum in (3.3) with $\hat{t}$ replaced by $t$. Consequently, when one measures the q-number $f(\omega\hat{t})$, the outcome of that measurement will be one of its eigenvalues $f(\omega t) \in \mathrm{Sp}\big(f(\omega\hat{t})\big)$. Thus, one can account for the appearance of a c-number function $f(\omega t)$ by referring only to the q-number function $f(\omega\hat{t})$ and its eigenvalues.

---

[10] A function of the form (3.3) is automatically periodic in the sense that it is invariant under the shift $\hat{t} \to \hat{t} + \frac{2\pi}{\omega}\hat{1}$.

## 3.4 Q-number derivatives

$f(\omega\hat{t})$ belongs to a possible state of the clock, and since the clock has so far not been assumed to be subject to any dynamical laws, there is no rate of change of $f(\omega\hat{t})$. Yet, an eigenvalue $f(\omega t)$ of $f(\omega\hat{t})$ does change relative to $t \in \mathrm{Sp}(\hat{t})$, implying that a derivative of $f(\omega t)$ is physical meaningful – such a derivative represents the instantaneous rate of change of $f(\omega t)$ with respect to clock time. Because the eigenvalues of $f(\omega\hat{t})$ have well-defined and physically meaningful derivatives, might it be similarly possible to define a derivative of $f(\omega\hat{t})$ with respect to $\hat{t}$, even in the absence of dynamics? In a simplified way, this problem may be summarised as follows:

$$\text{if} \quad \lambda \longrightarrow \hat{t}, \quad \text{then} \quad \frac{d}{d\lambda} \longrightarrow ? \tag{3.5}$$

Here $\lambda$ represents c-number time, and the arrow '$\longrightarrow$' should be understood to mean 'is replaced by'.[11]

To solve the problem described in (3.5), I define a derivative of $f(\omega\hat{t})$ with respect to $\hat{t}$ as follows:

$$\frac{df(\omega\hat{t})}{d\hat{t}} \stackrel{\text{def}}{=} \int_{-\infty}^{\infty} \frac{df(\omega t)}{dt}\hat{\Pi}_t(\hat{t})dt. \tag{3.6}$$

[11] This problem is related to the problem of formulating quantum theory without referring to classical concepts, as discussed by Deutsch (1984).

Here the c-number $f(\omega t)$ is an eigenvalues of $f(\omega\hat{t})$. Since $f(\omega\hat{t})$ is, by definition, equal to the expansion shown in (3.3), an alternative expression for (3.6) is

$$\frac{df(\omega\hat{t})}{d\hat{t}} = \sum_{n=-N}^{N} ia_n n\omega e^{in\omega\hat{t}}. \tag{3.7}$$

The expressions (3.7) and (3.6) are equivalent since the eigenvalues of (3.7) are exactly the $\frac{df(\omega t)}{dt}$ that appear in (3.6). Notably, the coefficients $b_n \stackrel{\text{def}}{=} (ia_n n\omega)$ in (3.7) have the property that $b_n{}^\dagger = b_{-n}$ for all $n$, so the derivative of a Hermitian q-number function is also Hermitian.

It is here that we find a remarkable connection between the algebra of the clock's descriptors and the q-number derivative. For example, the q-number derivative of $e^{i\omega\hat{t}}$ is

$$\frac{d\left(e^{i\omega\hat{t}}\right)}{d\hat{t}} = i\omega e^{i\omega\hat{t}}, \tag{3.8}$$

and because of the commutation relation (2.2), the derivative has an alternative formulation in terms of the following commutator[12]

$$i\left[\hat{h}, e^{i\omega\hat{t}}\right] = i\omega e^{i\omega\hat{t}}. \tag{3.9}$$

Hence, the q-number derivatives of $e^{i\omega\hat{t}}$ can be expressed algebraically in terms of the commutator (3.9). This is a striking and unique feature of the clock and its algebra: for if the commutation relation (2.2) were different, the correspondence between the q-number derivative (3.8) and the commutator (3.9) would not hold.

To elaborate on the correspondence between the q-number derivative and the commutation relation (3.9), consider the operator $i\left[\hat{h}, \bullet\right]$, where $\bullet$ denotes the operator's input. This operator maps q-numbers to q-numbers (*i.e.* it is a superoperator). Furthermore, for any q-numbers $\hat{A}$ and $\hat{B}$, the result of $i\left[\hat{h}, \bullet\right]$ on a linear combination of $\hat{A}$ and $\hat{B}$ is equal to the same linear combination of the superoperator's result on $\hat{A}$ and $\hat{B}$ separately. That is to say that, $i[\hat{h}, a\hat{A} + b\hat{B}] = ai[\hat{h}, \hat{A}] + bi[\hat{h}, \hat{B}]$, where $a$ and $b$ are real-valued constants. So, $i[\hat{h}, \bullet]$ is a linear operator, and under the action of this linear operator, the image of an arbitrary q-number function $f(\omega\hat{t})$ is

$$i[\hat{h}, f(\omega\hat{t})] = \sum_{n=-N}^{N} i a_n n \omega e^{in\omega\hat{t}}. \tag{3.10}$$

---

[12] The identity (3.9) can be derived as follows: because of the commutation relation (2.2), the unitary $e^{i\omega\hat{t}}$ is a translation operator of $\hat{h}$ so that $e^{-i\omega\hat{t}}\hat{h}e^{i\omega\hat{t}} = \hat{h} + \omega\hat{1}$, which can be rearranged to give the desired identity.

Manifestly, the right side of (3.10) is equivalent to the right side of (3.7) so that the q-number derivative has a more fundamental, algebraic expression:

$$\frac{df(\omega\hat{t})}{d\hat{t}} = i[\hat{h}, f(\omega\hat{t})]. \tag{3.11}$$

(3.11) is the pivotal result of this section: without referring to the c-number time $\lambda$, derivatives with respect to $\hat{t}$ can be defined algebraically – I will call this the q-number calculus. Remarkably, the q-number calculus is almost completely concealed in the Schrödinger picture due to that picture's focus on the c-number eigenvalues of descriptors. This obfuscates important questions that are natural to ask in the Heisenberg picture. For instance, one salient question is, can the q-number derivative defined in (3.11) be generalised to describe derivatives in a curved space-time? It is not obvious how to formulate that question in the Schrödinger picture.

# 4 Page & Wootters in the Heisenberg picture

I have shown how to formulate a q-number calculus in terms of the clock's descriptors and their algebraic properties. But how can the descriptors of other physical systems have equations of motion formulated in terms of $\hat{t}$? To address this question, consider a Page–Wootters universe that consists of a clock $\mathfrak{C}$ and a qubit $\mathfrak{Q}$.[13] To formulate the qubit's

[13] It is merely convenient to analyse a model universe consisting of a qubit and a clock. The model can be straightforwardly generalised to all finite systems – for example, the qubit can be replaced by a network of interacting qubits. The model can also account for systems with an infinite-dimensional

dynamics in terms of q-number time, let the qubit be represented by a triple of descriptors that are functions of $\hat{t}$

$$\hat{\boldsymbol{q}}(\omega\hat{t}) = \left(\hat{q}_x(\omega\hat{t}),\ \ \hat{q}_y(\omega\hat{t}),\ \ \hat{q}_z(\omega\hat{t})\right), \tag{4.1}$$

that satisfy the Pauli algebra

$$\left.\begin{aligned} \hat{q}_i(\omega\hat{t})\hat{q}_j(\omega\hat{t}) &= \delta_{ij}\hat{1} + i\epsilon_{ij}{}^{k}\hat{q}_k(\omega\hat{t}) \\ \left(\hat{q}_i(\omega\hat{t})\right)^{\dagger} &= \hat{q}_i(\omega\hat{t}) \end{aligned}\right\} \qquad (i, j, k \in \{x, y, z\}). \tag{4.2}$$

Here $\hat{1}$ is the unit observable of the composite system. It is not clear from (4.1) and (4.2) alone that $\hat{\boldsymbol{q}}(\omega\hat{t})$ is a function of $\hat{t}$**,** or anything of the sort, since all that is currently known about these descriptors is that they obey the Pauli algebra. Thus, let $\hat{\boldsymbol{q}}(\omega\hat{t})$ a sum of powers of $e^{i\omega\hat{t}}$ with q-number coefficients,

$$\hat{q}_j(\omega\hat{t}) \stackrel{\text{def}}{=} \sum_{n=-N}^{N} \hat{A}_{j,n} e^{in\omega\hat{t}} \qquad (j \in \{x, y, z\}), \tag{4.3}$$

where $N$ is an integer. I require that the coefficients $\{\hat{A}_{j,n}\}$ commute with $\hat{t}$ and $\hat{h}$ so that the products in the sum (4.3) are well-defined. Furthermore, the qubit's descriptors are Hermitian, and so $\hat{A}_{j,n}^{\dagger} = \hat{A}_{j,-n}$ for all $n \in \{-N, \dots, N\}$ and $j \in \{x, y, z\}$. Note that it is possible for the coefficients to be equal to zero for certain values of $n$ and $j$.

Because $\hat{\boldsymbol{q}}(\omega\hat{t})$ commutes with $\hat{t}$, it has a decomposition in terms of the $\{\hat{\Pi}_t(\hat{t}) : t \in \mathrm{Sp}(\hat{t})\}$,

$$\hat{\boldsymbol{q}}(\omega\hat{t}) = \int_{-\infty}^{\infty} \hat{\boldsymbol{q}}(\omega t)\hat{\Pi}_t(\hat{t})dt, \tag{4.4}$$

where $\hat{\boldsymbol{q}}(\omega t)$ is equivalent to sum in (4.3), but with $\hat{t}$ replaced by $t$. In this decomposition, the $\hat{\boldsymbol{q}}(\omega t)\hat{\Pi}_t(\hat{t})$ are the *relative descriptors* (Kuypers & Deutsch *op. cit.*), so named because these

Hilbert space. But since such systems are not necessarily periodic, a larger class of q-number functions would be required to analyse such systems. This larger class consists of polynomials and convergent series in $e^{i\omega\hat{t}}$.

descriptors represent the state of the qubit relative to the clock being in the state 'it is clock time $t$'.[14]

The relative descriptors satisfy the Pauli algebra: using (4.2) and the identity that $\hat{\boldsymbol{q}}(\omega\hat{t})\hat{\Pi}_t(\hat{t}) = \hat{\boldsymbol{q}}(\omega t)\hat{\Pi}_t(\hat{t})$ and $\hat{\Pi}_t(\hat{t})\hat{\Pi}_{t'}(\hat{t}) = \hat{\Pi}_t(\hat{t})\delta(t-t')$, where $\delta(t-t')$ denotes the Dirac delta function, one finds that

$$\left.\begin{aligned} \left(\hat{q}_i(\omega t)\hat{\Pi}_t(\hat{t})\right)\left(\hat{q}_j(\omega t)\hat{\Pi}_t(\hat{t})\right) &= \delta_{ij}\hat{\Pi}_t(\hat{t}) + i\epsilon_{ij}{}^{k}\hat{q}_k(\omega t)\hat{\Pi}_t(\hat{t}) \\ \left(\hat{q}_i(\omega t)\hat{\Pi}_t(\hat{t})\right)^{\dagger} &= \hat{q}_i(\omega t)\hat{\Pi}_t(\hat{t}) \end{aligned}\right\} \qquad (i,j,k \in \{x,y,z\}).$$

Here $\hat{\Pi}_t(\hat{t})$ is the 'relative unit observable' relative to the clock state 'it is time $t$'. Because the relative descriptors adhere to the Pauli algebra, they represent a fully fledged qubit. In the Page–Wootters construction, the relative descriptors represent the observable states of the qubit.

Since the q-number coefficients in (4.3) commute with $\hat{t}$ and $\hat{h}$, a derivative of $\hat{\boldsymbol{q}}(\omega\hat{t})$ with respect to $\hat{t}$ can be defined exactly like in (3.6), namely

$$\frac{d\hat{\boldsymbol{q}}(\omega\hat{t})}{d\hat{t}} \stackrel{\text{def}}{=} \int_{-\infty}^{\infty} \frac{d\hat{\boldsymbol{q}}(\omega t)}{dt}\hat{\Pi}_t(\hat{t})dt. \tag{4.5}$$

And this derivative has a more fundamental algebraic expression

$$\frac{d\hat{\boldsymbol{q}}(\omega\hat{t})}{d\hat{t}} = i\left[\hat{h}, \hat{\boldsymbol{q}}(\omega\hat{t})\right]. \tag{4.6}$$

Note that the right side of (4.6) is not necessarily zero because, due to $\hat{\boldsymbol{q}}(\omega\hat{t})$'s dependence on $\hat{t}$, it does not necessarily commute with $\hat{h}$. Incidentally, although $\hat{h}$ still satisfies the algebraic relation (2.2), it cannot represent the Hamiltonian of the clock, as the clock and qubit are space-like separated and should have mutually commuting descriptors – in §4.2 I will discuss this issue in more detail.

[14] These are the relative descriptors relative to the clock states 'it is clock time $t$' since a projector $\hat{\Pi}_t(\hat{t})$ projects onto a Heisenberg state for which $\hat{t}$ is sharp with value $t$ (see §3.2).

To describe the qubit's dynamics in terms of $\hat{t}$, the qubit's descriptors should have an equation of motion in terms of the derivative (4.6). Hence, I impose the constraint that the qubit's descriptors satisfy the Heisenberg equation of motion with respect to $\hat{t}$; this constraint can be formulated algebraically as follows:

$$[\hat{h} - \hat{H}, \hat{\boldsymbol{q}}(\omega\hat{t})] = 0 \;\; \Leftrightarrow \;\; \frac{d\hat{\boldsymbol{q}}(\omega\hat{t})}{d\hat{t}} = i[\hat{H}, \hat{\boldsymbol{q}}(\omega\hat{t})], \tag{4.7}$$

where $\hat{H}$ is the qubit's Hamiltonian. (4.7) demonstrates the central result of this paper: one can formulate the equations of motion of the qubit in terms of q-number time, which makes the c-number time unnecessary for formulating dynamics.

Remarkably, because of (4.7), the the qubit's relative descriptors automatically satisfy the Heisenberg equations of motion with respect to clock time: by using (4.4), (4.5), and (4.7), one finds that

$$\int_{-\infty}^{\infty} \left( \frac{d\hat{\boldsymbol{q}}(\omega t)}{dt} - [\hat{H}, \hat{\boldsymbol{q}}(\omega t)] \right) \hat{\Pi}_t(\hat{t}) dt = 0,$$

so

$$\frac{d\hat{\boldsymbol{q}}(\omega t)}{dt} = i[\hat{H}, \hat{\boldsymbol{q}}(\omega t)] \qquad\qquad \big(\text{for all } t \in \mathrm{Sp}(\hat{t})\big). \tag{4.8}$$

So because of (4.8), the qubit's relative descriptors (which represent the observable states of the qubit) obey the Heisenberg equation of motion with respect to clock time.

As an example of a solution to (4.7), consider the case in which the relative descriptors at clock time $t = 0$ are equal to the Pauli matrices, *i.e.* $\hat{\boldsymbol{q}}(0) = (\hat{\sigma}_x,\ \hat{\sigma}_y,\ \hat{\sigma}_z)$, and the qubit's Hamiltonian is $\hat{H} = \frac{1}{2}\omega\hat{\sigma}_x$. Then, an explicit solution to (4.7) is

$$\hat{\boldsymbol{q}}(\omega\hat{t}) = \big(\hat{\sigma}_x,\ \ \hat{\sigma}_y \cos(\omega\hat{t}) + \hat{\sigma}_z \sin(\omega\hat{t}),\ \ \hat{\sigma}_z \cos(\omega\hat{t}) - \hat{\sigma}_y \sin(\omega\hat{t})\big).$$

Hence, the qubit's dynamics can be formulated in terms of q-number time. Incidentally, note that the characteristic frequency $\omega$ is determined by the qubit's Hamiltonian.

## 4.1 Removing the c-number time

The Page–Wootters universe is an isolated quantum system, and therefore an arbitrary descriptor $\hat{A}(\lambda)$ of this model universe must satisfy the Heisenberg equation of motion with respect to c-number time $\lambda$, *i.e.*

$$\frac{d\hat{A}(\lambda)}{d\lambda} = i\big[\hat{\mathcal{H}}, \hat{A}(\lambda)\big]. \tag{4.9}$$

where $\hat{\mathcal{H}}$ is the Hamiltonian of the universe. Consequently, the arbitrary descriptor can be expressed as $\hat{A}(\lambda) = e^{i\lambda\hat{\mathcal{H}}}\hat{A}(0)e^{-i\lambda\hat{\mathcal{H}}}$, where $\hat{A}(0)$ is the descriptor at $\lambda = 0$. For instance, the triple $e^{i\hat{\mathcal{H}}\lambda}\hat{\boldsymbol{q}}(\omega\hat{t})e^{-i\hat{\mathcal{H}}\lambda}$ represents the $\lambda$-dependent descriptors of the qubit, and it adheres to both (4.9) and (4.7). So although the equation of motion of $e^{i\hat{\mathcal{H}}\lambda}\hat{\boldsymbol{q}}(\omega\hat{t})e^{-i\hat{\mathcal{H}}\lambda}$ has a formulation in terms in terms of q-number time, those descriptors also depend on the unphysical $\lambda$, which therefore still plays an explanatory role.

To obviate this problem, let $\hat{\mathcal{H}} = \hat{h}$: for this choice of $\hat{\mathcal{H}}$, the equation of motion (4.9) reduces to the requirement that, for any descriptor $\hat{A}(\lambda)$ of the model universe,

$$\frac{d\hat{A}(\lambda)}{d\lambda} = \frac{d\hat{A}(\lambda)}{d\hat{t}}. \tag{4.10}$$

In other words, due to (4.10), the derivative with respect to the unphysical c-number time can be replaced by derivatives with respect to the manifestly physical q-number time. To satisfy the constraint (4.10), the descriptors of the model universe are $e^{i\hat{\mathcal{H}}\lambda}\hat{t}e^{-i\hat{\mathcal{H}}\lambda} = \hat{t} + \lambda\hat{1}$ and $e^{i\hat{\mathcal{H}}\lambda}\hat{\boldsymbol{q}}(\omega\hat{t})e^{-i\hat{\mathcal{H}}\lambda} = \hat{\boldsymbol{q}}\left(\omega\left(\hat{t} + \lambda\hat{1}\right)\right)$, while all other descriptors of the model are completely invariant with respect to $\lambda$. Additionally, because $\hat{\mathcal{H}} = \hat{h}$, the qubit's relative descriptors commute with $\hat{\mathcal{H}}$ so that the relative descriptors are valid observables of the model, as required.

Finally, to avoid the descriptors' expectation values from depending on $\lambda$, let the Heisenberg state, denoted $|\Psi\rangle$, be an eigenstate of $\hat{\mathcal{H}}$. Consequently, the expectation value of an arbitrary

descriptor $\hat{A}(\lambda)$ is independent of $\lambda$, *i.e.* $\langle\Psi|\hat{A}(\lambda)|\Psi\rangle = \langle\Psi|e^{i\lambda\hat{\mathcal{H}}}\hat{A}(0)e^{-i\lambda\hat{\mathcal{H}}}|\Psi\rangle = \langle\Psi|\hat{A}(0)|\Psi\rangle$ for all $\lambda$. As a result of these constraints, c-number time does not play an explanatory role in the Page–Wootters construction: no expectation value of the model depends on it; any derivative with respect to $\lambda$ can be substituted for a derivative with respect to $\hat{t}$; and a system's equations of motion can instead be formulated entirely in terms of $\hat{t}$.

## 4.2 Moving between stationary pictures

With the Heisenberg-picture Page–Wootters construction completed, it seems timely to reflect on the connections between the Schrödinger and Heisenberg picture variants of the construction. I have summarised these connections in the chart below.

| **The Schrödinger picture** | | **The Heisenberg picture** |
|---|---|---|
| $\lvert\Psi_S\rangle = \int_{-\infty}^{\infty} \lvert\psi(t)\rangle\lvert t\rangle dt$ | $\xleftrightarrow{\text{absolute state}}$ | $\hat{\boldsymbol{q}}(\omega\hat{t}) = \int_{-\infty}^{\infty} \hat{\boldsymbol{q}}(\omega t)\hat{\Pi}_t(\hat{t})dt$ |
| $\lvert\psi(t)\rangle\lvert t\rangle$ | $\xleftrightarrow{\text{relative state}}$ | $\hat{\boldsymbol{q}}(\omega t)\hat{\Pi}_t(\hat{t})$ |
| $\hat{\mathcal{H}}_S = \hat{H}_S + \hat{h}_S$ | $\xleftrightarrow{\text{Hamiltonian}}$ | $\hat{\mathcal{H}} = \hat{h}$ |
| $i\frac{d}{dt}\lvert\psi(t)\rangle = \hat{H}_S\lvert\psi(t)\rangle$ | $\xleftrightarrow{\text{equation of motion}}$ | $\frac{d\hat{\boldsymbol{q}}(\omega\hat{t})}{d\hat{t}} = i[\hat{H}, \hat{\boldsymbol{q}}(\omega\hat{t})]$ |

Here the arrows '↔' establish qualitative connections between objects on the left and right columns in the chart.

Moreover, the Schrödinger- and Heisenberg-pciture Page–Wootters constructions are mathematically identical: as Rijavec (2022) has demonstrated, if one starts with a Schrödinger-picture Page–Wootters model that consists of a qubit and clock, then by using the unitary transformation $V \stackrel{\text{def}}{=} e^{-i\hat{H}_S\hat{t}_S}$, one obtains the Heisenberg-picture construction as follows:

$$\left.\begin{array}{lll} V^\dagger\hat{h}_S V = \hat{h}_S - \hat{H}_S, & V^\dagger\hat{t}_S V = \hat{t}_S, & V^\dagger\hat{\boldsymbol{q}}(0)V = e^{i\hat{H}_S\hat{t}_S}\hat{\boldsymbol{q}}(0)e^{-i\hat{H}_S\hat{t}_S} \\ V^\dagger\hat{\mathcal{H}}_S V = \hat{h}_S, & V^\dagger\hat{H}_S V = \hat{H}_S, & V^\dagger|\Psi_S\rangle = |\psi(0)\rangle|TL\rangle \end{array}\right\}, \tag{4.11}$$

where $|TL\rangle \stackrel{\text{def}}{=} \int_{-\infty}^{\infty} |t\rangle dt$ is the so-called 'time line' state (Giovannetti *et al.* 2015). Similarly, the inverse transformation $V^{\dagger}$ maps the Heisenberg-picture to the Schrödinger-picture construction.[15] As is evident from (4.11), the Heisenberg descriptors of the clock and the qubit are a mixture of their Schrödinger picture counterparts – for instance, the Hamiltonian of the clock in the Heisenberg picture is equal to $\hat{h}_S - \hat{H}_S$. (Note that $\hat{h}_S - \hat{H}_S$ must commute with the qubit's triple of descriptors, which results in the equation of motion (4.7).)

Although the Schrödinger and the Heisenberg picture are empirically identical, they are conceptually different: when the Page–Wootters construction is expressed in the Schrödinger picture, one need only refer to the eigenvalues of the q-number time. Yet in the Heisenberg picture, we must fundamentally reinterpret what functions of time, and derivatives with respect to time, are.

## 4.3 Missed but remembered

Because the Heisenberg state is an eigenstate of $\hat{h}$, the expectation value of each relative descriptor is non-zero – *i.e.* $\langle \hat{\Pi}_t(\hat{t}) \rangle > 0$ for all $t \in \mathrm{Sp}(\hat{t})$. Remarkably, this feature is not necessary for the Page–Wootters construction to explain time and dynamics (Marletto & Vedral 2017). For instance, no observer could detect that the expectation value $\langle \hat{\Pi}_t(\hat{t}) \rangle$ is identically zero for some interval of $\hat{t}$'s eigenvalues because, even if certain clock times are 'missing' from the Page–Wootters universe, every system in that universe would nonetheless end up in exactly the state they would have been in if those 'missing' clock times had been there. This is analogous to a situation in the novel *Permutation City* (Greg Egan 1994), in which a simulated person, Paul, is being rendered by a computer at half-second intervals and not at intermediate moments:

[15] This transformation also allows us to incorporate time-dependent Hamiltonians into the formalism presented here. For instance, if the Hamiltonian of the qubit contains explicit time-dependence, denoted as $\hat{H}(\hat{t})$, then the transformation that maps the Schrödinger picture to the Heisenberg picture will be a time-ordered exponential of $\hat{H}(\hat{t})$ – see for instance Giovannetti *et al.* (*op. cit.*).

*'Paul counted – and the truth was, he felt no different … And that made sense [because] every half-second, his brain was ending up in exactly the state it would have been in if nothing had been left out.'*

## 5 Open problems in the Page–Wootters construction

The clock in the Page–Wootters universe is ideally accurate: it has infinitely many distinguishable states (*i.e.* it has infinitely many eigenstates), and so there is no limit on how accurately it can keep track of time. However, this also implies that one can store an infinite amount of information in the clock (Deutsch 2004), despite it being widely believed that only a finite amount of information can be stored in any compact region of space due to Bekenstein's bound (1973). Thus, the universe should contain a multitude of unideal clocks, which must be finite physical systems.[16]

That this menagerie of clocks can give rise to something like a global time at all should be due to the clocks being synchronised (Wootters *op. cit.*). For instance, consider two quantum clocks $\mathfrak{C}_1$ and $\mathfrak{C}_2$, with respective pairs of canonically conjugate descriptors $(\hat{t}_1, \hat{h}_1)$, and $(\hat{t}_2, \hat{h}_2)$, where the descriptors of $\mathfrak{C}_1$ commute with those of $\mathfrak{C}_2$.[17] The clock $\mathfrak{C}_2$ is perfectly synchronised with $\mathfrak{C}_1$ if the observable $\hat{t}_1 - \hat{t}_2$ is sharp with value zero because, in that case, if one measures $\hat{t}_1$ and $\hat{t}_2$ separately and subtracts the two measurement results, zero offset will be detected between the clocks.

The above-described patchwork of clocks bears striking similarities to the atlas of a manifold. An atlas consists of a family of charts (functions from the manifold to $\mathbb{R}^n$). Firstly, a chart resembles an individual clock in that an ideal clock, like $\mathfrak{C}_1$, provides a function from $\hat{t}_1$ to the continuum through its spectrum: $\mathrm{Sp}(\hat{t}_1) = \mathbb{R}$. Furthermore, the eigenvalues $\mathrm{Sp}(\hat{t}_1)$ represent clock time, much like how the chart represents coordinates on the manifold. Secondly, the requirement that $\hat{t}_1 - \hat{t}_2$ is sharp with value zero allows us to determine how clock time $\hat{t}_1$ is related to clock time $\hat{t}_2$. This is similar to how coordinates of one chart can be related to the coordinates of another via a transition map (*i.e.* a homeomorphism between two charts in a region where those charts overlap).

[16] An unideal clock must have a time observable with finitely many eigenvalues, so such an observable represents discrete time. In that case, one should be able formulate the equations of motion of the rest of the universe as difference equations, where for comparatively large systems, these difference equations will be approximately identical to the Heisenberg equation (4.7). I shall leave the explicit formulation of a Heisenberg Page–Wootters construction with unideal clocks for future research. For more on how to use finite physical systems as clocks, see the work of Smith & Ahmadi (2019) and Rijavec (*op. cit.*).

[17] Notably, $\mathfrak{C}_1$ and $\mathfrak{C}_2$ are ideal clocks because they have canonical conjugate observables that necessarily live on an infinite dimensional Hilbert space. Although such ideal clocks cannot be physical, studying them will nonetheless be informative since it will tell us how well two physical clocks can be synchronised in the absence of other limitations.

Hence, I conjecture that the properties of a space-time manifold can be formulated in terms of q-numbers. Formulating space-time manifolds in this way *prima facie* requires an extension of the Page–Wootters model – for instance, the model should incorporate space (Singh 2020). Furthermore, in both special and general relativity, the structure of space-time imposes severe restrictions on the behaviour of clocks: absolute simultaneity is not permitted by special and general relativity, which implies that two space-like separated clocks cannot be synchronised in all reference frames. Likewise, due to the effects of time dilation, clocks will stop 'ticking' in lockstep when they start moving relative to one another. In a more general Page–Wootters construction, it seems that clocks such as $\mathfrak{C}_1$ and $\mathfrak{C}_2$ should adhere to such restrictions. I shall leave these question to be addressed in future research.

## Acknowledgements

I am grateful for the many discussions with David Deutsch, Chiara Marletto, Simone Rijavec, and Charles Alexandre Bédard – they helped me understand the material in this paper better than I would have been able to on my own. I am also grateful for the comments on earlier drafts of this paper provided by David Deutsch, Chiara Marletto, Simone Rijavec, Charles Alexandre Bédard, Anicet Tibau Vidal, Eric Marcus, Liberty Fitz-Claridge, and two anonymous referees. This work was supported in part by the Prins Bernhard Cultuurfonds.